\documentclass[doublecol]{arXiv} 

\usepackage{dcolumn}    
\usepackage{bm} 
\usepackage{graphicx}
\usepackage{amsmath}    
\usepackage{latexsym}
\usepackage{amsfonts}   
\usepackage{amssymb}
\usepackage{array}      
\usepackage{epsfig}
\usepackage{txfonts}
\usepackage[colorlinks=true,linkcolor=blue,urlcolor=magenta,citecolor=magenta]{hyperref}
\usepackage{color}
\usepackage{hyperref}
\usepackage[normalem]{ulem}

\newcommand{\ket}[1]{\left\vert#1\right\rangle}
\newcommand{\bra}[1]{\left\langle#1\right\vert}
\newcommand{\ketbra}[2]{|#1\rangle\langle#2|}
\newcommand{\braket}[2]{\langle#1|#2\rangle}

\newcommand{\bla}{bla\\bla\\bla\\bla\\bla}

\begin{document}
\title{Quantum work statistics of controlled evolutions}

\author{Steve Campbell$^{1,2,3}$}

\institute{$^1$School of Physics, University College Dublin, Belfield, Dublin 4, Ireland \\
$^2$Centre for Quantum Engineering, Science, and Technology, University College Dublin, Belfield, Dublin 4, Ireland \\
$^3$Dahlem Center for Complex Quantum Systems, Freie Universit\"at Berlin, Arnimallee 14, 14195 Berlin, Germany}       

\abstract{We use the quantum work statistics to characterize the controlled dynamics governed by a counterdiabatic driving field. Focusing on the Shannon entropy of the work probability distribution, $P(W)$, we demonstrate that the thermodynamics of a controlled evolution serves as an insightful tool for studying the non-equilibrium dynamics of complex quantum systems. In particular, we show that the entropy of $P(W)$ recovers the expected scaling according to the Kibble-Zurek mechanism for the Landau-Zener model. Furthermore, we propose that the entropy of the work distribution provides a useful summary statistic for characterizing the need and complexity of the control fields for many-body systems.}
\date{\today}
\maketitle

\section{Introduction} 
From studies of fundamental phenomena to the development of practical quantum devices, achieving breakthroughs in virtually all areas requires advanced techniques to manipulate complex systems. The need for coherent control is ubiquitous. This demand has precipitated the development of several techniques to coherently engineer the dynamics of quantum systems~\cite{STAreview, OCreview} and their experimental demonstration~\cite{Bason2011, Zhang2013, Lam2021, Zhou2020, Zhang2018}. Shortcuts-to-adiabaticity, which are particularly elegant approaches stemming from the adiabatic theorem, have been the focus of sustained recent interest~\cite{STAreview}. These approaches consider the quantum evolution from the outset and reverse engineer the dynamics to attain a specific target and therefore, in principle, do not require any complex optimization to be performed~\cite{Demirplak, Berry}, although such optimization can significantly enhance their efficacy and implementability~\cite{Ieva2023}. 

While the obvious motivation, from a practical stance, is to design control protocols to achieve a particular aim, it is intuitive that understanding the requirements to render a system controllable will provide a deeper understanding of the system at hand~\cite{RabitzReview}. In the case of a shortcut-to-adiabaticity, since at least the start and end points coincide with the adiabatic states, this intuition implies that assessing the control protocols themselves can provide a versatile tool to examine the complex non-equilibrium dynamics being suppressed~\cite{Demirplak2, Cost2016, Puebla2020, CalzettaPRA, TakahasiPRR2020, SalazarPRE, Santos2015, TongPRA2022, Campbell2017, Funo2017}.

We follow this ethos by assessing the work statistics associated with controlled evolutions generated by a counterdiabatic field, complementing previous studies that examined the work distribution, with a particular focus on its fluctuations as a useful indicator for the thermodynamic cost of control~\cite{Funo2017}. As a key figure of merit we use the entropy of the distribution~\cite{Kiely2023}, which provides a useful summary tool characterizing the work statistics and, therefore, goes beyond the typically considered first and second moments. We demonstrate that the thermodynamics of a controlled evolution can be used to examine the non-equilibrium dynamics being suppressed and also provides a further means to characterize the complexity of the control.

\section{Quantum work statistics}
Defining work in a quantum mechanically consistent way has proven to be a subtle issue~\cite{QThermoBook}. We will consider the two-point measurement approach~\cite{TPM}, where the work done is a stochastic variable given by the probability distribution
\begin{equation}
\label{PW}
    P(W) = \sum\limits_{n,m} ~p_n p_{m|n} \delta\left[(W - (E_m(t) - E_n(0)\right]\;,
\end{equation}
where $p_n$ is the probability of getting an initial outcome $n$ and $p_{m|n}$ is the subsequent conditional probability of recording outcome $m$ from the second energy measurement. This approach to defining work has proven to be very useful, allowing to recover and extend results from classical stochastic thermodynamics and allowing to examine the highly nontrivial role that quantum coherence can play~\cite{QThermoBook}. We will use Eq.~\eqref{PW} to examine the characteristics of a controlled evolution that achieves an adiabatic dynamics in a finite time. This is particularly interesting since, naively, if we only look at the start and end of the protocol, by construction, the work performed will be precisely the adiabatic work, implying that the control was thermodynamically free. However, several works have demonstrated that this view is not entirely accurate and that there is an unavoidable thermodynamic cost associated with achieving control~\cite{Demirplak2, Cost2016, CalzettaPRA, SalazarPRE, Santos2015, Campbell2017, Funo2017}.

We will focus on arguably the simplest protocol to achieve a finite time adiabatic dynamics through counterdiabatic~\cite{Demirplak} or transitionless~\cite{Berry} driving. Consider the Hamiltonian given by its spectral decomposition
\begin{equation}
    \mathcal{H}_0(t)=\sum_n \epsilon_n(t)\ket{\phi_n(t)}\bra{\phi_n(t)}.
\end{equation}
We can achieve a perfect adiabatic dynamics by adding an additional control term to suppress the non-equilibrium excitations generated due to the finiteness of the drive~\cite{Demirplak, Berry}
\begin{equation}
\label{eq:cdcd}
    \mathcal{H}_{1}(t)=i\sum_n \Big( \ketbra{\dot{\phi_n}}{\phi_n} 
    -\braket{\phi_n}{\dot{\phi_n}}\ketbra{\phi_n}{\phi_n} \Big),
\end{equation}
where for brevity we have dropped the explicit time dependence and here, and throughout, we assume units such that $\hbar=1$. In what follows, we will assume that the system begins in an eigenstate, specifically the ground state, $\ket{\psi(0)}=\ket{\phi_0(0)}$, of $\mathcal{H}_0$. To compute the work distribution we need the evolution of the state after the initial measurement, which for an arbitrary process requires to explicitly simulate the correct time-ordered dynamics. However, by construction the controlled evolution precisely tracks the adiabatic ground state, i.e. $\ket{\psi(t)}=U(t)\ket{\psi(0)}=\ket{\phi_0(t)}$, and therefore the calculation of $p_{m|n}$ is greatly simplified being given by
\begin{equation}
\label{Pmn}
p_{m|n} = \vert \langle \Phi_m(t)\ket{\phi_n(t)} \vert^2.
\end{equation}
where $\ket{\Phi_m(t)}$ are the eigenstates of the full generator, i.e. $\mathcal{H}_0 + \mathcal{H}_1$ and clearly also $E_m$ entering Eq.~\eqref{PW} are the corresponding energy eigenvalues. To ensure the protocol realizes an adiabatic dynamics with respect to the bare Hamiltonian, the condition $\mathcal{H}_1=0$ is enforced at the start and end points of the drive. Therefore, at these points the work distribution will be precisely that of an adiabatic protocol and thus provide no further insight. However, examining the properties of Eq.~\eqref{PW} {\it during} the controlled evolution can reveal clear signatures of the non-equilibrium dynamics being suppressed. Remarkably, however, the average of the work distribution, $\langle W \rangle$, during the control protocol is precisely the same as the adiabatic work~\cite{Funo2017}. This establishes that not all moments of the distribution provide insight into the thermodynamics of the control protocol. 

In order to capture the richness of the full distribution, we focus on recently proposed summary statistic to characterize $P(W)$ by computing its Shannon entropy
\begin{equation}\label{HW}
    H_W = - \sum_W P(W) \ln P(W).
\end{equation}
As shown in Ref.~\cite{Kiely2023}, $H_W$ can be bounded by two distinct contributions, one stemming from the entropy of the initial state and a second directly related to the coherence generated by the driving protocol. Thus, due to our assumption that the system begins in an eigenstate of $\mathcal{H}_0$, the entropy of $P(W)$ for the controlled dynamics can be related purely to the coherence generated in the eigenbasis of the generator~\cite{Kiely2023}.  

With these tools we now show that the work distribution, and in particular the entropy $H_W$, provides an insightful metric for characterizing the thermodynamics of control and that the properties of $P(W)$ allows to explore, in a very general manner, the dynamical response of the system to arbitrary drives. We first demonstrate that the entropy of the controlled work distribution admits a scaling in line with the predictions of the Kibble-Zurek mechanism~\cite{Kibble, Zurek}. This establishes that the controlled evolution provides a versatile means to study generic non-equilibrium dynamics and complements recent studies on the thermodynamics of traversing a quantum phase transition~\cite{Deffner2017, Esposito2020, AdolfoPRL2018, delCampoPRR2020, Krissia2020, Zawadski2023}. As a second application, we propose to use $H_W$ as a proxy for the complexity of the control by assessing the work distribution arising from different implementations of counterdiabatic driving for many-body systems. Specifically we compare the full counterdiabatic term, Eq.~\eqref{eq:cdcd}, with a control field which is tailored for only a specific eigenstate, explicitly showing that the latter, arguably more simple control Hamiltonian, generally results in a smaller $H_W$.

\section{Kibble-Zurek scaling}
We begin by considering a two-level driven system as captured by the Landau-Zener (LZ) model~\footnote{While we follow the usual naming convention for Eq.~\eqref{LZmodel}, it has been noted in several works that a more accurate name is Landau–Zener–St\"uckelberg–Majorana model~\cite{Nori2023}.}
\begin{equation}
\label{LZmodel}
 \mathcal{H}_{\rm LZ}(t)=\frac{\Delta}{2}\sigma_x+ \frac{g(t)}{2} \sigma_z,
\end{equation} 
where $\sigma_i$ are the Pauli matrices. We fix the time dependent part of Eq.~\eqref{LZmodel} to have the form $g(t)= g_0 + g_d \left(\frac{t}{\tau_Q}\right)$, i.e. a linear ramp, although remark that our results are qualitatively unaffected for other choices of ramping profile. Despite its simplicity, the LZ model captures a remarkably versatile range of physical phenomena~\cite{Nori2023} and will serve to demonstrate the key insights achievable by examining the thermodynamics of control. In particular, due to the avoided crossing in the energy spectrum of Eq.~\eqref{LZmodel}, the LZ model is known to be the minimal model capturing the dynamics of a quantum phase transition according to the Kibble-Zurek mechanism~\cite{Damski2005, Damski2006}. Several works have explored the dynamics of control for systems described by Eq.~\eqref{LZmodel}, including using information about the transition between adiabatic and impulsive regimes to tailor more suitable control protocols~\cite{Carolan2022}.

\begin{figure}[t]
\centering
(a)\\
\includegraphics[width=0.75\columnwidth]{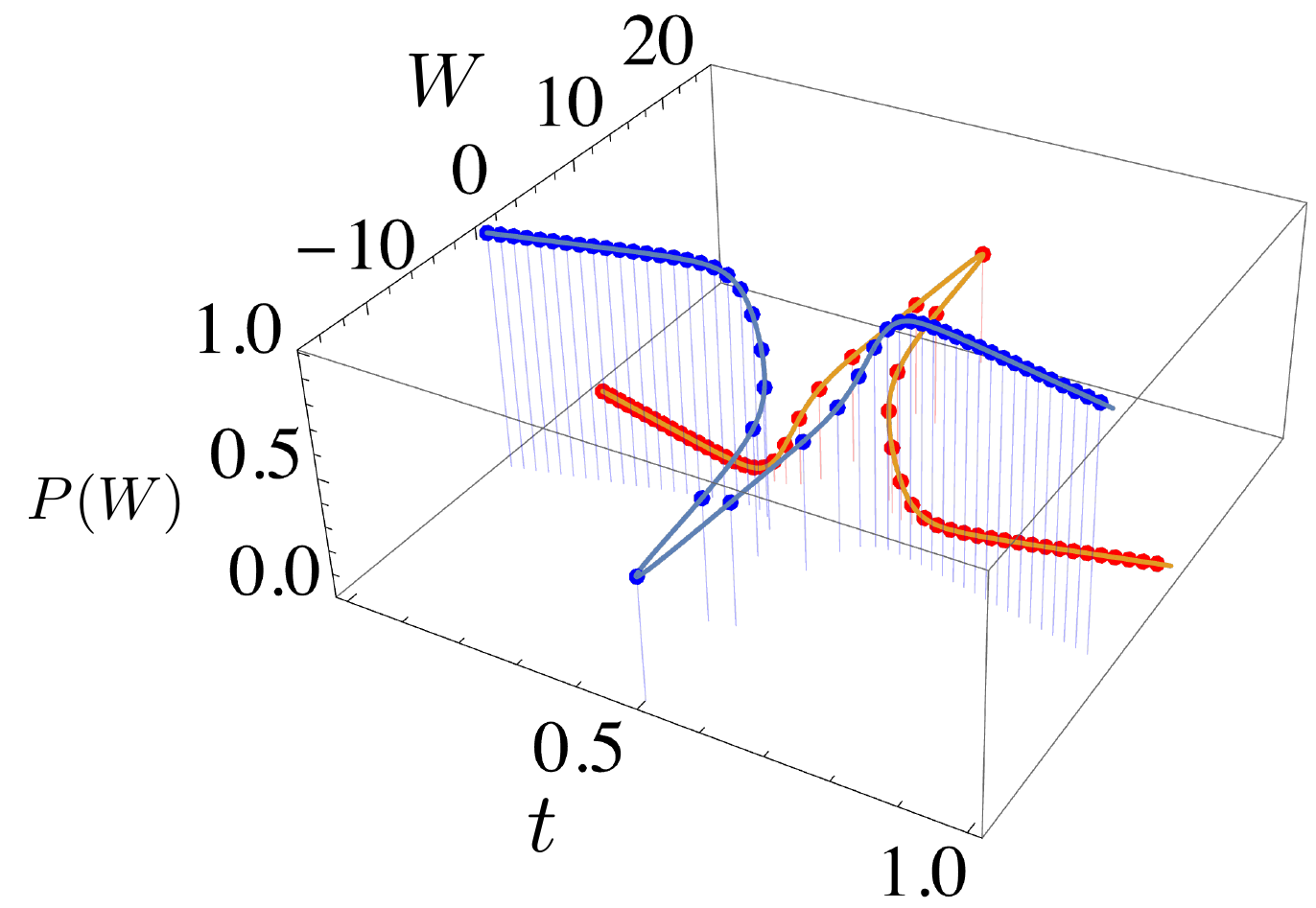}\\
(b)\\
\includegraphics[width=0.75\columnwidth]{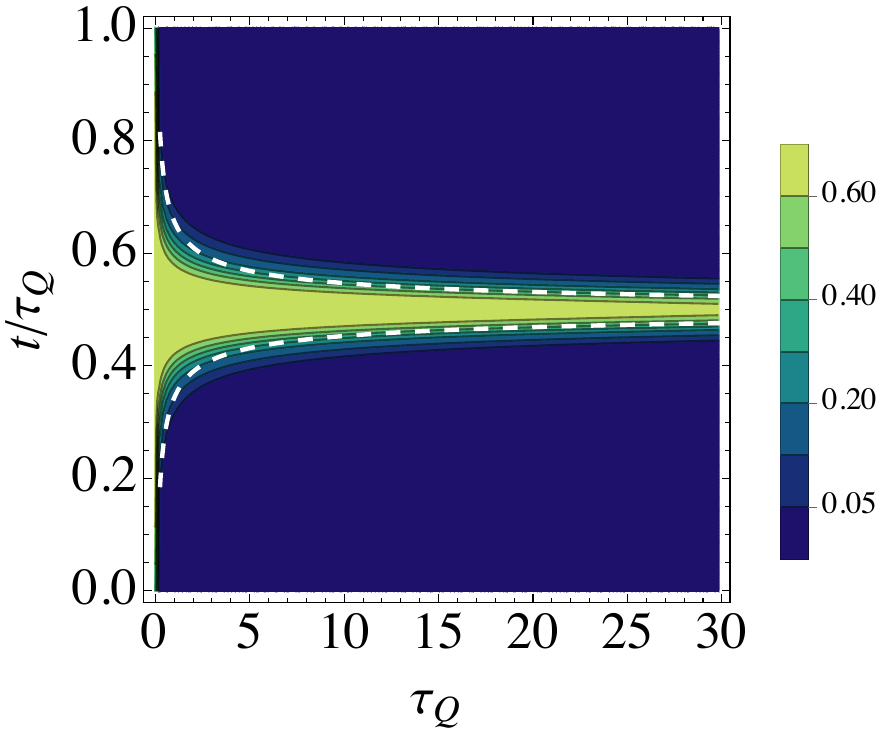}\\
(c)\\
\includegraphics[width=0.75\columnwidth]{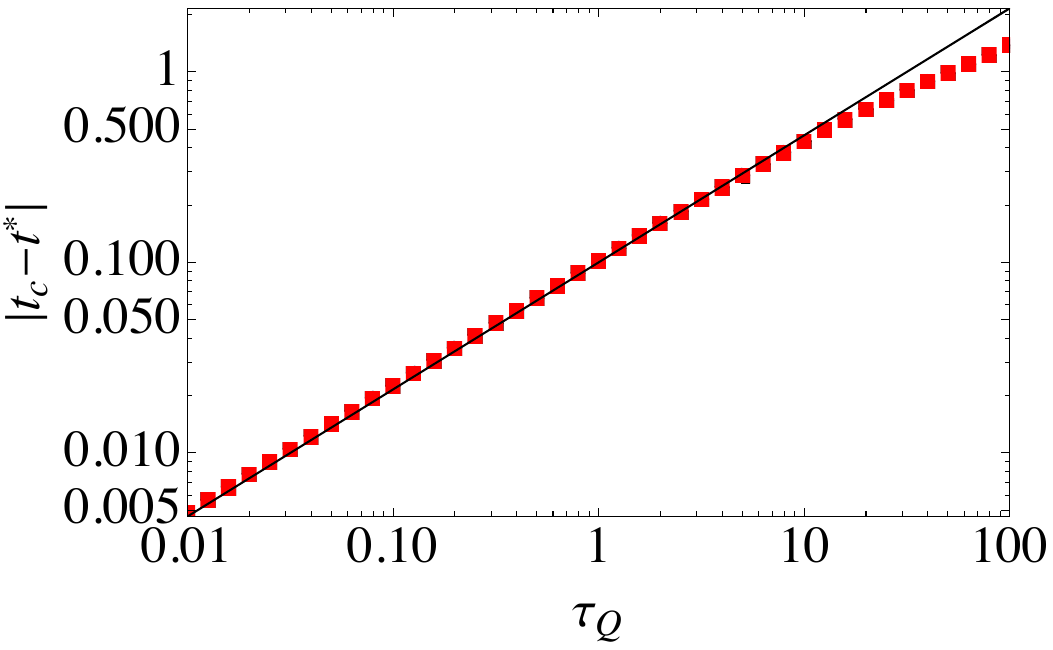}
\caption{(a) Work distribution for the LZ model. Darker, blue points are the ground state and lighter, red points are for the excited state. Solid lines correspond to the analytic values. We fix the quench duration $\tau_Q=1$. (b) Entropy of the distribution, $H_W$, versus total protocol duration. White dashed lines delineate the transition between adiabatic and impulse regimes from usual Kibble-Zurek arguments~\cite{Damski2005, Carolan2022}. (c) Size of the impulse regime showing the scaling of $\tau_Q^{2/3}$ in line with predictions~\cite{Puebla2020}. Here $t_c$ is the time at which the protocol reaches the avoided crossing, i.e. $\tau_Q/2$, and $t^*$ is the point where the rate of change of the entropy is maximum.  All plots assume a linear ramp with $g_0=-10$, $g_d=20$, $\Delta=0.5$.}
\label{fig1}
\end{figure}

In Fig.~\ref{fig1}(a) we show the explicit work distribution, $P(W)$ for the controlled evolution, where the points are numerically evaluated, but due to the simplicity of the model $P(W)$ can be readily determined analytically, shown with the continuous solid lines. We clearly see a change in behavior around the avoided crossing where the distribution becomes double peaked, reflecting the fact that need-for and cost-of control maximize in this region~\cite{Funo2017, Campbell2017}. It is important to stress that $\langle W \rangle$ for this distribution precisely coincides with the adiabatic work~\cite{Funo2017}. This highlights that the first moment of the distribution provides no useful insights, however, from Fig.~\ref{fig1}(a) it is evident that this will not be the case for higher order moments.

We can employ Eq.~\eqref{HW} to more carefully examine the characteristics of the distribution, particularly in the region in which $P(W)$ becomes more complex due to the counterdiabatic driving field. In Fig.~\ref{fig1}(b) we show the entropy of $P(W)$ as a function of quench duration where we can clearly see a transition between two regions. The darker, blue region corresponds to distributions with almost zero entropy, indicating that the distribution is $\delta$-peaked, i.e. only a single eigenstate of the dynamical Hamiltonian is playing a role. However, a significant change in the entropy of the distribution occurs as the system approaches the avoided crossing. Here we see the entropy quickly increases and saturates to its maximal value of $\ln 2$. This transition is clearly reminiscent of the adiabatic-impulse approximation~\cite{Zurek}. The superposed dashed, white lines correspond to the analytical expressions for the adiabatic-impulse crossover times that can be directly computed by the usual Kibble-Zurek arguments~\cite{Damski2005, Carolan2022} and shows good agreement with the characteristics of $H_W$. 

Rather than the heuristic arguments typically employed to determine the scaling behavior, we can use $H_W$ to estimate the the crossover between the adiabatic and impulse regimes as the time, $t^*$, when the rate of change of $H_W$ is maximal. By direct computation we find that the value of $t^*$ is similar to the crossover times delineated by the white-dashed lines in Fig.~\ref{fig1}(b). Extending the ramp duration, $\tau_Q$, to much larger values allows to examine how the extent of the impulse regime scales with the quench duration. Fig.~\ref{fig1}(c) shows the size of the impulse regime, $\vert t_c - t^* \vert$, where $t_c$ is the time at which the ramp hits the avoided crossing point, i.e. $g(t_c)=0$, with the solid line establishing that the scaling is proportional to $\tau_Q^{2/3}$, which is in excellent agreement with the Kibble-Zurek predictions~\footnote{Notice that the LZ model is shares the essential features of the quantum Ising model which exhibits a characteristic $\tau_Q^{1/2}$ scaling. As discussed in Ref.~\cite{Puebla2020}, however, the bare LZ model admits a scaling with exponent $2/3$.}.

We remark that signatures of Kibble-Zurek scaling have been reported in the moments of the work distribution~\cite{Esposito2020}. However, our analysis is distinct since it focuses explicitly on the controlled evolution where, by construction, the dynamics is perfectly adiabatic with regards to the bare Hamiltonian. Our results highlight that the non-equilibrium dynamics can be studied by examining the controlled evolution, however, this necessitates looking at higher order moments of the work distribution~\cite{Funo2017}. While we have focused on counterdiabatic driving, it is relevant to ask whether other approaches that transiently leave the adiabatic manifold but still achieve the same adiabatic start and end points also allows to quantitatively study the non-equilibrium dynamics being suppressed. While a careful analysis on a technique by technique basis would be required, we have verified that a similar behavior to the results reported here can be achieved for local counterdiabatic driving~\cite{AdolfoPRL2013}. Our analysis indicates that, while counterdiabatic driving may suffer limitations associated with the difficulty of its implementation for all but the simplest of systems, it nevertheless provides a useful and versatile tool for exploring generic non-equilibrium dynamics.

\section{Complexity of the control}
We next assess how the entropy of $P(W)$ can be used as a proxy for the complexity of the applied control. Equation~\eqref{eq:cdcd} ensures no transitions between {\it any} eigenstates occur. However, in most circumstances not every state will need to be driven in a transitionless manner~\cite{Cost2016, Demirplak2}. In particular, if the system is initialised in an eigenstate of $\mathcal{H}_0$ then only transitions from this state need to be suppressed. Therefore, we can achieve the same level of control by employing a control Hamiltonian tailored to the specific state via 
\begin{equation}
\label{restrictedcontrol}
\mathcal{H}_{1}^{n}(t) = i \left[ \frac{d}{dt} \ketbra{\phi_n}{\phi_n}, \ketbra{\phi_n}{\phi_n} \right].
\end{equation}
It follows from similar arguments outlined in Ref.~\cite{Funo2017} that the average work of the state evolved according to $\mathcal{H}_0+\mathcal{H}_{1}^{n}$ still coincides exactly with the adiabatic work. However, this Hamiltonian is, at least in some sense, ``simpler" than Eq.~\eqref{eq:cdcd} since it only suppresses transitions away from a specific eigenstate and in general requires a lower energetic overhead to implement~\cite{Cost2016}. In what follows, we demonstrate that $H_W$ allows to make this intuition more quantitative.

\begin{figure}[t]
\centering
(a)\\
\includegraphics[width=0.75\columnwidth]{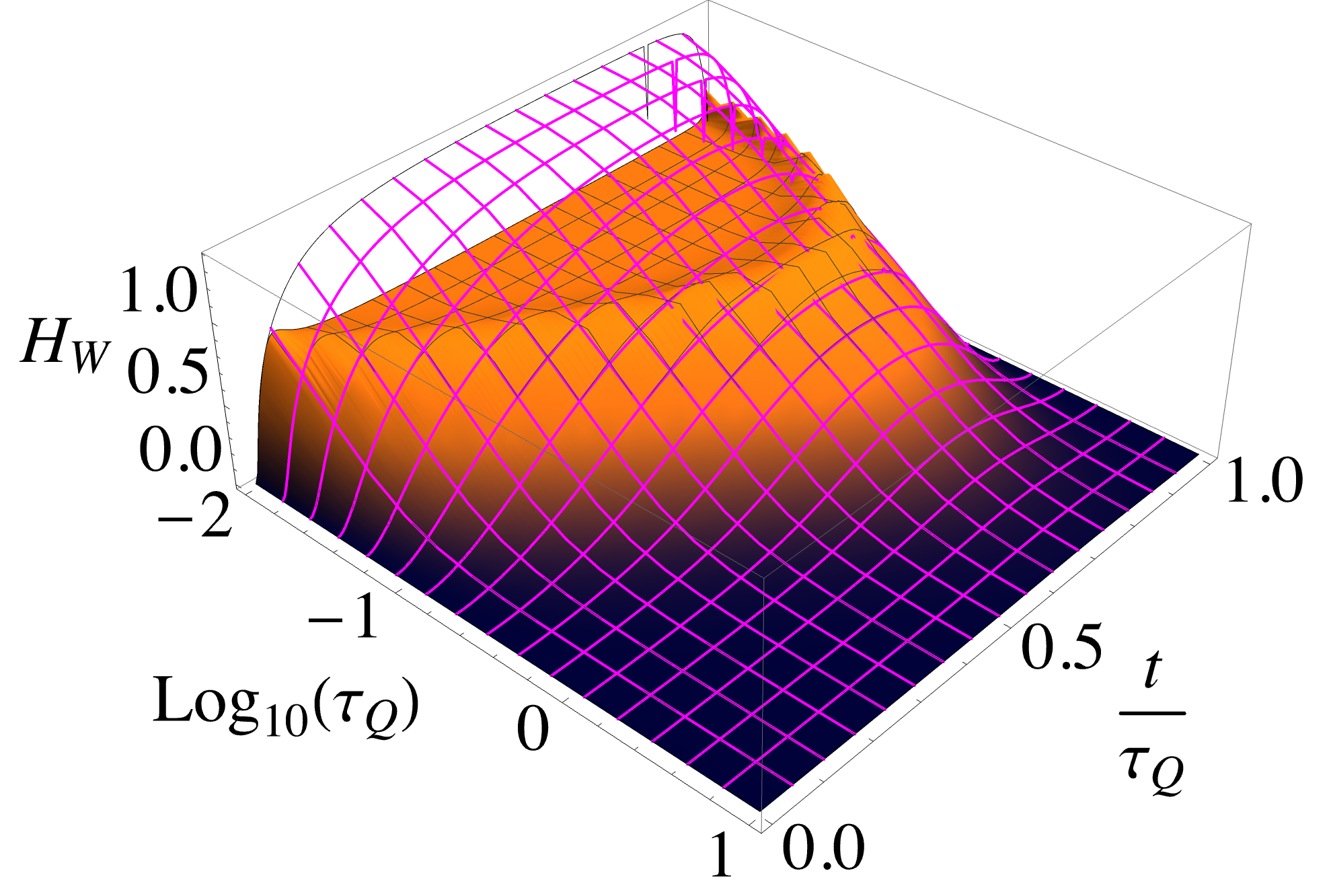}\\
(b)\\
\includegraphics[width=0.75\columnwidth]{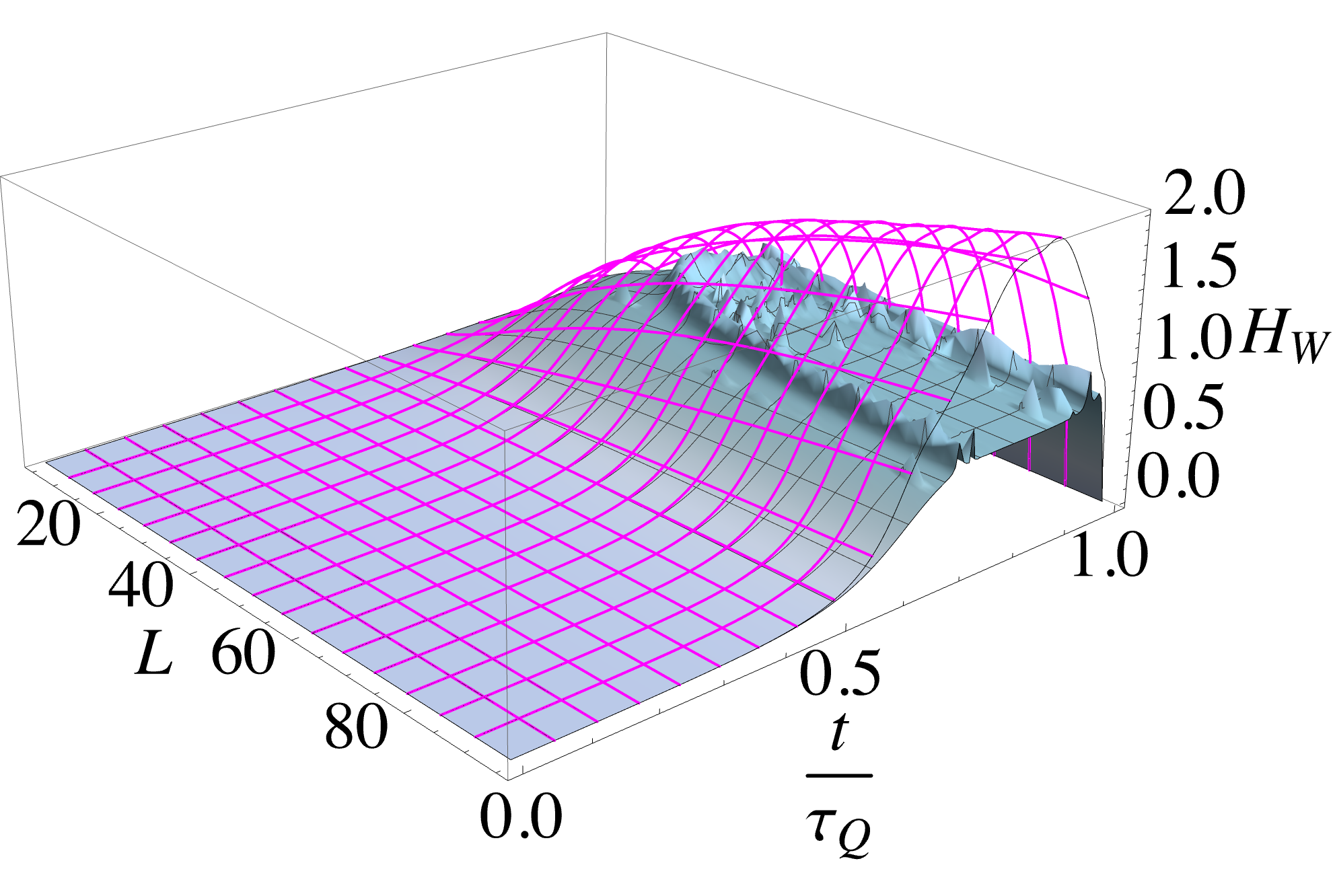}
\caption{(a) Entropy of the distribution, $H_W$, for the Ising model for a fixed number of spins, $L=5$, as a function of quench duration. (b) Entropy of the distribution, $H_W$, for the LMG model for a fixed quench duration, $\tau_Q=1$, for various sized systems. Both panels assume the smooth sinusodial ramp, Eq.~\eqref{sineramp}, with $g_0=2$ and $g_d=-1.2$. The transparent mesh corresponds to the full CD driving term, Eq.~\eqref{eq:cdcd}, while the solid planes are when the control is achieved using Eq.~\eqref{restrictedcontrol}.}
\label{fig2}
\end{figure}
We consider the Ising, $\mathcal{H}_I$, and the Lipkin-Meshkov-Glick (LMG), $\mathcal{H}_{LMG}$, models which correspond to many-body systems with different interaction ranges
\begin{eqnarray}
    &\mathcal{H}_I&= -\sum_{i=1}^{L}\left[g(t) \sigma_i^z - \sigma_i^x\sigma_{i+1}^x\right], \\
    &\mathcal{H}_{LMG}&=-\sum_{i=1}^{L}g(t) \sigma_i^z-\frac{1}{L}\sum_{i<j}\sigma_i^x\sigma_{j}^x,
\end{eqnarray}
where for the former we assume periodic boundary conditions. In the thermodynamic limit both models host a quantum phase transition at $g(t)=1$ and we set
\begin{equation}
\label{sineramp}
g(t) = g_0 + g_d \sin^2\left( \frac{\pi t}{2 \tau_Q} \right).
\end{equation}

In Fig.~\ref{fig2} we compare the entropy of the work distributions for the two realizations of a controlled evolution associated with implementing $\mathcal{H}_1$ or $\mathcal{H}_1^n$. In panel (a), we consider the Ising model of fixed size $L=5$ and examine how the entropy, and therefore the complexity, of the distribution, changes for various quench durations. The transparent mesh shows $H_W$ when the control is achieved using the full counterdiabatic Hamiltonian, Eq.~\eqref{eq:cdcd}, while the solid plane corresponds to employing Eq.~\eqref{restrictedcontrol}. Evidently, for long quench durations approaching the adiabatic limit the entropy of the distributions coincide as $H_W\to0$. While both distributions have the same average value regardless of the speed of the protocol, fast quenches demonstrate the significant difference between the two. We find $H_W$ is larger when the full $\mathcal{H}_1$ is employed, indicating that the work distribution is more complex for this driving protocol. In contrast, for control achieved via $\mathcal{H}_1^n$, the entropy of the distribution saturates to a value $\simeq\!\!\ln 2$ during the dynamics when the control term is active. Panel (b) demonstrates that a qualitatively similar behavior is exhibited when the protocol duration, $\tau_Q$, is fixed and the system size varies. Here, we examine the LMG model where for the ramp given by Eq.~\eqref{sineramp} the ground state energy gap closes as the system is driven. Consequently the need for control techniques to suppress unwanted transitions increases~\cite{Cost2016} and we see the entropy of the controlled work distribution arising from $\mathcal{H}_1$ grows with system size. Computing $H_W$ corresponding to $\mathcal{H}_1^n$, however, we find it is consistently lower and again tends to saturate close to $\simeq\!\!\ln 2$, indicating a simpler distribution. This establishes that the thermodynamics of control, and in particular the entropy of the work probability distribution, is a useful summary tool to capture and characterize differences in protocols that achieve the same evolution.

\section{Conclusions}
We have examined the quantum work distribution for controlled evolutions governed by counterdiabatic Hamiltonians. Since, by construction, the generator gives rise to a perfectly adiabatic dynamics, determining the probability distribution becomes particularly simple and serves to provide useful insight, both in terms of the non-equilibrium dynamics being suppressed and the resource intensiveness of the control. We demonstrated this through two exemplary settings: regarding the former, we showed that non-equilibrium dynamics captured by the Kibble-Zurek mechanism can be readily studied by examining the work statistics of the controlled evolution. For the latter, we showed that the work probability distribution allows to define an {\it ad hoc} notion of complexity. Through explicit examples of the quantum Ising and LMG models, we showed that distinct Hamiltonians that nevertheless achieve the same evolution can have significantly different thermodynamic behaviors. In both settings, the Shannon entropy of the work distribution, $H_W$, served as the key summary statistic. By focusing on an initial state that is an eigenstate of the initial Hamiltonian, $H_W$ can be directly related to the coherences generated by the controlled drive through the framework of Ref.~\cite{Kiely2023} and, therefore, our approach may provide an alternative means to quantify the cost of control by directly relating it to the thermodynamic cost of creating coherence~\cite{Huber}. It would be interesting to examine how our results are affected for systems initialized in mixed states, e.g. thermal states where the inclusion of a finite temperature will naturally lead to an increase in the entropy of the work distribution. However, since the populations are conserved when undergoing counterdiabatic driving we expect that the entropy will admit a splitting between two distinct contributions, one arising from the initial mixedness of the state and the other due to the coherence generated by the control protocol, in line with the results from Ref.~\cite{Kiely2023}. Finally we remark that our results can also be applied to classical systems undergoing counterdiabatic driving~\cite{DeffnerPRX, BravettiPRE} where the work statistics will again provide information regarding complexity of the control.  Our results further strengthens the motivation to use control as more than a means-to-an-end, establishing that the controlled dynamics provides an insightful tool to study the equilibrium and non-equilibrium properties of complex systems. 

\acknowledgments
This work was supported by the Science Foundation Ireland Starting Investigator Research Grant No. 18/SIRG/5508 and the Alexander von Humboldt Foundation.

\end{document}